\documentclass[aps,prd,twocolumn,showpacs,floatfix,nofootinbib,superscriptaddress]{revtex4-1}
\usepackage{amsmath}
\usepackage{amssymb}
\usepackage{graphicx}
\usepackage{xurl}
\usepackage{hyperref}
\usepackage{color}
\usepackage{mathtools}
\usepackage{times}
\usepackage{bm}
\usepackage{xcolor}

\usepackage[normalem]{ulem}

\newcommand{\Dquad}{\qquad\qquad}

\newcommand{\nnn}{\nonumber \\}

\newcommand{\beeq}{\begin{equation}}
	\newcommand{\eneq}{\end{equation}}
\newcommand{\bear}{\begin{eqnarray}}
	\newcommand{\enar}{\end{eqnarray}}

\newcommand{\LL}{\mathcal{L}}    

\newcommand{\CC}{\mathcal{C}}    
\newcommand{\rbar}{\bar r}       

\newcommand{\cc}{\lambda}        

\newcommand{\HH}{\mathcal{H}}   

\newcommand{\OO}{\mathcal{O}}

\newcommand{\al}{\alpha}
\newcommand{\be}{\beta}
\newcommand{\ga}{\gamma}
\newcommand{\de}{\delta}

\newcommand{\pa}{\partial}

\newcommand{\RR}{\mathcal{R}}  

\begin{document}

\title{Infrared Sensitivity of Cosmological Probes in Horndeski Theory}

\author{Matteo Magi}
\email{matteo.magi@uzh.ch}
\affiliation{Center for Theoretical Astrophysics and Cosmology,
Department of Astrophysics,
University of Zurich, Winterthurerstrasse 190,
CH-8057, Zurich, Switzerland}
\author{Jaiyul Yoo}
\email{jyoo@physik.uzh.ch}
\affiliation{Center for Theoretical Astrophysics and Cosmology,
Department of Astrophysics,
University of Zurich, Winterthurerstrasse 190,
CH-8057, Zurich, Switzerland}
\affiliation{Physics Institute, University of Zurich,
Winterthurerstrasse 190, CH-8057, Zurich, Switzerland}

\date{\today}

\begin{abstract}

Cosmological probes constructed in large-scale surveys are independent of the underlying theory of gravity, and their relativistic descriptions are indeed applicable to any theory of gravity. It was shown that the presence of fluctuations with wavelength much larger than the characteristic scales of the surveys has no impact on cosmological probes, if the matter content is adiabatic and the Einstein equations are used. In this paper we study the sensitivity of cosmological probes to infrared fluctuations in Horndeski theory. We find that the extra degree of freedom in the Horndeski scalar field can induce sensitivity to infrared fluctuations in the cosmological probes, even when the matter components are adiabatic on large scales. A generalized adiabatic condition including the extra dof, in contrast, guarantees that cosmological probes are devoid of infrared sensitivity, and this solution corresponds to the adiabatic modes à la Weinberg in Horndeski theory, which can be removed by a coordinate transformation in the infrared limit. We discuss the implications of our findings and the connections to the initial conditions.

\end{abstract}

\maketitle

	\section{Introduction}	
	
The standard model of cosmology has been extremely successful in describing cosmological observations from small scales to very large scales. The key assumption in the standard model is that the general theory of relativity correctly describes gravitational interactions on large scales, and this assumption is well tested, thanks to the remarkably precise measurements in the solar system~\cite{willConfrontationGeneralRelativity2014,schlammingerTestEquivalencePrinciple2008,williamsProgressLunarLaser2004} and recent detections of gravitational waves~\cite{abbottGravitationalWavesGammarays2017, abbottGW151226ObservationGravitational2016, abbottGW170817ObservationGravitational2017, abbottObservationGravitationalWaves2016a, goldsteinOrdinaryShortGammaRay2017,abbottTestsGeneralRelativity2016,bertiTestingGeneralRelativity2015}. However, on cosmological scales where gravity is weak, precision measurements are rather difficult, though observations of cosmic microwave background (CMB) anisotropies provide tight constraints on any deviation from general relativity in the early universe~\cite{Planck:2015bue}. In particular, the cosmic acceleration in the late time and the existence of elusive dark matter in the standard model are often invoked for the possibility that gravitational interactions might be different on large scales in the late time (see, e.g.,~\cite{copelandDynamicsDarkEnergy2006a,peeblesCosmologicalConstantDark2003,fengDarkMatterCandidates2010,tanabashiReviewParticlePhysics2018,weinbergObservationalProbesCosmic2013} for recent reviews).

From a theoretical point of view, Lovelock theorem~\cite{lovelockEinsteinTensorIts1971} states that general relativity is \textit{unique} in the sense that it is the only local theory that leads to equations of motion that contain up to second derivative of the four-dimensional spacetime metric.
Once the premise of the Lovelock theorem is dropped, a plethora of alternatives are permitted. The most well-known modification is the Brans-Dicke theory~\cite{bransMachPrincipleRelativistic1961}, in which an extra scalar field couples to the metric tensor, acting as an additional gravitational degree of freedom. The Brans-Dicke scalar traces the matter content of the universe, following Mach principle, and it effectively plays the role of a spacetime-dependent gravitational coupling, different from the Newton constant. This type of scalar-tensor theory of gravity was further generalized by Horndeski~\cite{horndeskiSecondorderScalartensorField1974} to classify all the possibilities with one extra scalar field in four spacetime dimensions that give rise to {\it no} more than second-order derivatives in the equations of motion. Theories with higher-order derivatives are often plagued with particles with negative kinetic energy leading to Ostrogradski instabilities, and they are not favored (see, e.g.,~\cite{achourDegenerateHigherOrder2016, crisostomiExtendedScalarTensorTheories2016, gleyzesNewClassConsistent2015, langloisDegenerateHigherDerivative2016} for higher derivative theories without ghosts).

Recently, great attention was paid to Horndeski theory as an alternative to dark energy, in which a modification of gravity on large scales is responsible for the late-time cosmic acceleration~\cite{cliftonModifiedGravityCosmology2012,koyamaCosmologicalTestsModified2016,heisenbergSystematicApproachGeneralisations2019,kobayashiHorndeskiTheoryReview2019,beltranjimenezStabilityHorndeskiVectortensor2013,belliniConstraintsDeviationsLCDM2016,Bamba:2012cp,Nojiri:2010wj,Nojiri:2017ncd}. The theory is further constrained by the recent observation of a neutron star merger with a black hole through gravitational-wave propagation~\cite{abbottGravitationalWavesGammarays2017, abbottGW170817ObservationGravitational2017} but, despite this development, Horndeski theory still encompasses rich phenomenology~\cite{cliftonModifiedGravityCosmology2012,heisenbergSystematicApproachGeneralisations2019,koyamaCosmologicalTestsModified2016,creminelliDarkEnergyInstabilitiesInduced2020,defeliceTheories2010,gleyzesHealthyTheoriesHorndeski2015,gleyzesEssentialBuildingBlocks2013,charmousisGeneralSecondOrder2012,sotiriouTheoriesGravity2010,copelandDynamicsDarkEnergy2006a,heisenbergHorndeskiQuantumLoupe2020}.
Here we study Horndeski theory as a model of gravity beyond general relativity.

The theoretical descriptions of cosmological probes such as the CMB temperature anisotropies, galaxy clustering, luminosity distance, and weak gravitational lensing are independent of the theory of gravity, be it general relativity or a metric theory such as Horndeski theory. In fact, under the assumption that light propagates along null geodesics of the metric, a geometric construction of cosmological probes is possible, which leads to the gauge-invariant relativistic descriptions of cosmological probes (see~\cite{sasakiMagnitudeRedshiftRelation1987,futamaseLightPropagationDistance1989,bonvinFluctuationsLuminosityDistance2006,yooNewPerspectiveGalaxy2009,yooGeneralRelativisticDescription2010,challinorLinearPowerSpectrum2011a,bonvinWhatGalaxySurveys2011a,jeongLargescaleClusteringGalaxies2012,ben-dayanValueH_0Inhomogeneous2014,umehNonlinearRelativisticCorrections2014,yooLinearOrderRelativisticEffect2014,bertaccaObservedGalaxyNumber2014,didioGalaxyNumberCounts2014,bonvinWeCareDistance2015,koyamaObservedGalaxyBispectrum2018,magiSecondorderGaugeinvariantFormalism2022}). While Newtonian contributions are of course dominant on small scales, the relativistic contributions in cosmological probes take over on large scales. The presence of fluctuations on scales larger than the horizon scale, therefore, indicates that cosmological probes may be strongly affected by such large-scale (infrared) modes.

This observation contradicts our intuition that while cosmological probes are modulated by infrared fluctuations, their influence progressively wanes as the wavelength of the fluctuations increases. Detailed studies using the relativistic description of cosmological probes~\cite{jeongLargescaleClusteringGalaxies2012,scaccabarozziGalaxyTwoPointCorrelation2018,grimmGalaxyPowerSpectrum2020,castorinaObservedGalaxyPower2022,biernCorrelationFunctionLuminosity2017,biernGaugeInvarianceInfraredDivergences2017,baumgartnerMonopoleFluctuationCMB2021,mitsouInfraredSensitivityRelativistic2023} show that while infrared fluctuations can in principle lead to pathological behavior of cosmological probes, such infrared contributions are in fact canceled to confirm our intuition, if the matter content is adiabatic and the Einstein equations are used. In this paper, we attempt to answer the question of whether general relativity is special in terms of this infrared sensitivity in cosmological probes by investigating large-scale solutions in Horndeski theory under the assumption that the matter content is adiabatic.
	
In Section~\ref{uno} we outline the theoretical framework underlying this paper. We first recapitulate the results of our previous work on general conditions for infrared insensitivity, and then present our strategy for applying these results to Horndeski theory.
In Section~\ref{due}, we search for infrared solutions in Horndeski theory assuming adiabatic conditions between the matter fields on large scales. First, we derive the solution in the Brans-Dicke limit and then infer the solution in the Horndeski theory. In Section~\ref{tre} we demonstrate the connections of the infrared solutions in Horndeski theory to the adiabatic modes by Weinberg. In Section~\ref{quattro} we conclude and discuss our results.

	\section{Theoretical Framework}\label{uno}

In this Section, we introduce the concept of infrared sensitivity of cosmological probes, as presented in a previous paper~\cite{magiConditionsAbsenceInfrared2023a}. After summarizing the key results that characterize the infrared sensitivity, we review the Horndeski theory of gravity, which we will use as a gravitational theory beyond general relativity.

	\subsection{General conditions for infrared insensitivity in any gravity theories}\label{ir}
 
Here we briefly summarize the main findings in~\cite{magiConditionsAbsenceInfrared2023a} for the impact of infrared fluctuations on cosmological probes.
Consider a perturbed Friedmann-Lemaître-Robertson-Walker (FLRW) spacetime with line element given by
    \beeq\label{FRWpert}
        \begin{split}
        ds^2=&-a^2(1 + 2\al)d\eta^2-2a^2\pa_i\be d\eta dx^i
        \\&
        +a^2\left[( 1+2\varphi)\de_{ij}+2\pa_i\pa_j\ga\right] dx^i dx^j\,,    
        \end{split}
    \eneq
where~$a(\eta)$ is the scale factor as a function of conformal time,~$x^i$ are comoving coordinates, and $\al(\eta,\bm x)$, $\be(\eta,\bm{x})$, $\ga(\eta,\bm x)$, and $\varphi(\eta,\bm x)$ are the scalar metric perturbations expressed in a generic gauge.
We have assumed spatially flat slices ($K=0$), and throughout the paper we will restrict our attention to linear-order scalar perturbations.
We specify the matter content of the universe as a perfect fluid consisting of a mixture of pressure-less matter, denoted by the subscript~$_m$, and radiation, denoted by the subscript~$_\ga$. The energy-momentum tensor of such perfect fluid reads
    \beeq\label{energymomentum}
        \begin{split}
        -T^0{}_0=\rho=\rho_m+\rho_\ga \,,\qquad\quad  T^i{}_j=p\de^i_j= p_\ga\de^i_j\,,
        \\
        -T^0{}_i=\left(\bar p+\bar\rho\right)\pa_i v=\bar\rho_m\pa_i v_m+\frac43\bar\rho_\ga\pa_i v_\ga\,,\quad
        \end{split}
    \eneq
where~$\rho$,~$p$ and~$v$ are the total energy density, pressure, and velocity potential of the fluid respectively. Quantities with a subscript refer to the individual components, and we denote background FLRW fluid quantities with an overbar.

In~\cite{magiConditionsAbsenceInfrared2023a} we have shown that if the fluid quantities above satisfy certain relations, then cosmological probes of large-scale structure and CMB temperature anisotropies do not show a pronounced sensitivity to perturbations with wavelengths much larger than the distance between the observer and the source. 
In other words, the cosmological probes are affected by longer wavelength perturbations with progressively smaller impact, vanishing in the limit the wavelength becomes infinite.
We call such long-wavelength perturbations \textit{infrared} perturbations, and the conditions for infrared insensitivity in~\cite{magiConditionsAbsenceInfrared2023a} guarantee the absence of infrared sensitivity for cosmological probes, because contributions of such infrared fluctuations cancel each other.
Such conditions can be conveniently expressed by introducing the comoving-gauge curvature perturbations~$\RR$ and the uniform-density gauge curvature perturbation~$\zeta$. These curvature perturbations are gauge invariant at linear order in perturbations and are defined as
	\bear\label{defcurv}
	\RR_I:=\varphi-\HH v_I\,,\qquad\quad \zeta_I:=\varphi-\HH\frac{\de\rho_I}{\bar\rho'_I}\,,
	\enar
where the subscript~$_I$ indicates that the fluid quantities on the right-hand side are calculated for a single species, which in our case is either pressure-less matter ($I=m$) or radiation ($I=\ga)$. When no subscript is specified, the definitions in Eq.~($\ref{defcurv}$) apply to the total fluid.
The general conditions for the absence of infrared sensitivity in the scalar sector are derived in~\cite{magiConditionsAbsenceInfrared2023a} as
    \bear\label{condit}
    \frac{\de\rho_m}{\bar\rho'_m}=\frac{\de\rho_\ga}{\bar\rho'_\ga}\,,\qquad \RR_m=\zeta_m\,,\qquad \nabla v_m(\eta_z)=\nabla v_\ga(\eta_z)\,,\quad
    \enar
where the equality requirement holds only on sufficiently large scales and at all times, except for the relative velocity which only needs to vanish at the redshift of the source, corresponding to the conformal time~$\eta_z$.

Given a generic scalar perturbation~$f(x)$ in the theoretical description of a probe with characteristic comoving size~$\rbar$, e.g., the comoving distance of the last scattering surface~$\rbar_*$ for the CMB temperature anisotropies, we can define the long-wavelength contribution~$f_L(x)$ by smoothing~$f(x)$ in real space on a scale~$R\gg \rbar$.
Independently of the details of the smoothing,~$f_L(x)$ can be written in a spatial expansion such that the terms in the expansion with the lowest number of spatial derivatives dominate for an appropriate choice of sufficiently large~$R$. For a spatial expansion on the line of sight $x^i=\rbar n^i$ we derive the following expression for long-wavelength fluctuations
    \beeq\label{exp}
        \begin{split}
        f_L(\eta,\bm x)&=f_L(\eta)+x^i\pa_if_L(\eta)+\OO(\bm x^2)
        \\&
        =:f_0(\eta)+\rbar f_1(\eta,\bm n)+\OO(\bm x^2)\,.  
        \end{split}
    \eneq
In~\cite{magiConditionsAbsenceInfrared2023a} we showed that the infrared sensitivity is controlled by the first two coefficients of the above spatial expansion, hence, our conditions in Eq.~($\ref{condit}$) apply to the coefficients~$f_0$ and~$f_1$ of the perturbations in Eq.~($\ref{condit}$). When the general conditions in Eq.~($\ref{condit}$) are satisfied, we proved in~\cite{magiConditionsAbsenceInfrared2023a} that the first two coefficients
of the luminosity distance fluctuation~$\de D_L$~\cite{biernCorrelationFunctionLuminosity2017,biernGaugeInvarianceInfraredDivergences2017,yooUnifiedTreatmentLuminosity2016,scaccabarozziLightConeObservablesGaugeInvariance2017,yooMaximumCosmologicalInformation2020}, galaxy clustering~$\de_g$~\cite{jeongLargescaleClusteringGalaxies2012,grimmGalaxyPowerSpectrum2020,scaccabarozziGalaxyTwoPointCorrelation2018,castorinaObservedGalaxyPower2022,yooNonGaussianitySqueezedThreepoint2022}, and the CMB temperature anisotropies~$\Theta$~\cite{baumgartnerMonopoleFluctuationCMB2021,yooBackgroundPhotonTemperature2019} completely vanish
    \bear\label{IRins}
    \left(\de D_L\right)_{0,1}=0\,,\qquad \left(\de_g\right)_{0,1}=0 \,,\qquad  \Theta _{0,1}=0\,.
    \enar
We recall that~$\de D_L$ and~$\de_g$ are essentially the fluctuations of the physical area and volume occupied by the source, together with the matter density fluctuations at the source position, while~$\Theta$ is essentially the same as the observed redshift fluctuations, assuming that recombination takes place instantaneously at the temperature set by atomic physics.
Despite the presence of numerous relativistic contributions in the first two coefficients of the spatial expansion, the conditions for their precise cancellation in the cosmological probes~$\de D_L$,~$\de_g$, and~$\Theta$, are independent of the underlying theory of gravity (or valid for any gravity theories), and it only requires that light follows null geodesics of the metric in Eq.~($\ref{FRWpert}$) from the source to the position of the observer.

The first equality in Eq.~($\ref{condit}$) is the large-scale adiabatic condition for the matter content of the universe. In terms of total fluid quantities it reads
    \bear\label{adiab}
    \frac{\de p}{\bar p'}=\frac{\de\rho}{\bar\rho'}=\frac{\de\rho_m}{\bar\rho_m'}=\frac{\de\rho_\ga}{\bar\rho_\ga'}\,.
    \enar
Throughout this paper we assume that pressure-less matter and radiation satisfy the adiabaticity condition in Eq.~($\ref{adiab}$) on large scales, and we do not discuss the mechanism by which adiabaticity is established on such large scales. A direct consequence of adiabaticity is that~$\zeta$ is conserved in time on large scales:
    \bear
    \zeta=\zeta_m=\zeta_\ga\,,\qquad \zeta'=\frac1{3\left(\bar p+\bar\rho\right)^2}\left(\bar\rho'\de p-\bar p'\de\rho\right)=0\,,
    \enar
which originates from the covariant conservation of the energy-momentum tensor. All the equivalences hold for the first two coefficients of a spatial expansion, but we suppress the subscripts~$_0$ and~$_1$ and simply refer to them as being valid on large scales. Given the assumption that matter components are adiabatic, if the following condition for the total matter component holds
    \bear\label{RZ}
    \RR=\zeta\,,
    \enar
the general conditions for infrared insensitivity are fully satisfied. In fact, since pressure-less matter and radiation redshift differently, the above condition implies
    \bear
    \RR_m=\RR_\ga=\RR=\zeta=\zeta_m=\zeta_\ga\,, \qquad\quad v_m=v_\ga \,,
    \enar
which are the stronger version of Eq.~($\ref{condit}$). Again, these equalities hold only on large scales.

The equality of Eq.~($\ref{RZ}$) is a condition that should be checked in any theory of gravity. For general relativity, it is well-known that the condition~$\RR=\zeta$ is automatically satisfied by the Einstein equations if perturbations are adiabatic on large scales. Here we investigate whether~$\RR=\zeta$ also holds in Horndeski gravity theory, or in other words, whether adiabaticity in the matter content alone is sufficient to guarantee the infrared insensitivity of cosmological probes beyond the Einstein gravity.

	\subsection{Horndeski theory}\label{hornd}

The Horndeski theory of gravity is the most general scalar-tensor theory in four dimensions with up to second-order differential covariant equations of motion, which allows no degree of freedom for ghosts and thereby ensures the stability of the theory (see, e.g.,~\cite{cliftonModifiedGravityCosmology2012,koyamaCosmologicalTestsModified2016,heisenbergSystematicApproachGeneralisations2019,kobayashiHorndeskiTheoryReview2019} for recent reviews). We present the basic equations of the Horndeski theory of gravity and apply them to cosmology.

Horndeski theory is specified by the action~\cite{horndeskiSecondorderScalartensorField1974}
    \bear
    S_{H}=\int d^4x\sqrt{-g}\left(\LL_2+\LL_3+\LL_4+\LL_5\right)\,,
    \enar
and the individual Lagrangians are
    \bear\label{Li}
    \LL_2&=&G_2\,,\Dquad\LL_3=G_3\Box\phi\,,
    \nnn
    \LL_4&=&G_4R+\pa_XG_{4}\left[\left(\Box\phi\right)^2-\left(\nabla_\mu\nabla_\nu\phi\right)^2  \right]\,,
    \nnn
    \LL_5&=&-\frac16\pa_XG_{5}\big[ \left(\Box\phi\right)^3-            3\Box\phi\left(\nabla_\mu\nabla_\nu\phi\right)^2
    +2\left(\nabla_\mu\nabla_\nu\phi\right)^3 \big]
    \nnn
    &&+G_5G_{\mu\nu}\nabla^\mu\nabla^\nu\phi\,,
    \enar
where~$\phi(\eta,\bm x)$ is the additional scalar field that is responsible for gravity together with the metric tensor~$g_{\mu\nu}(\eta,\bm x)$. The functions~$G_i(\phi,X)$, with $i=2,3,4,5$, are arbitrary functions of~$\phi$ and its kinetic term $X=-\frac12g^{\mu\nu}\nabla_\mu\phi\nabla_\nu\phi$. Finally,~$G_{\mu\nu}$ is the Einstein tensor,~$R$ is the Ricci scalar, and $\Box=g^{\mu\nu}\nabla_\mu\nabla_\nu$ is the D'Alambert operator in curved spacetime. Horndeski~\cite{horndeskiSecondorderScalartensorField1974} proved that the sum of the above Lagrangians~$\LL_i$ keeps the field equations up to second-order derivatives, so that Ostrogradsky~\cite{ostrogradskyMemoiresEquationsDifferentielles1850} instabilities do not affect the theory.

The recent observation of a neutron star merger with a black hole~\cite{abbottGravitationalWavesGammarays2017, abbottGW170817ObservationGravitational2017} put tight constraints on the Lagrangians in Eq.~($\ref{Li}$) by the condition that gravitational waves propagate at the speed of light. Therefore, the observationally viable action for Horndeski theory reduces to~\cite{kreischCosmologicalConstraintsHorndeski2018a,creminelliDarkEnergyGW1708172017, lombriserBreakingDarkDegeneracy2016, saksteinImplicationsNeutronStar2017}
    \bear\label{horn}
    S_H=\int d^4x\sqrt{-g}\bigg(  G_4(\phi)R+G_3(\phi,X)\Box\phi +G_2(\phi,X)\bigg).\quad
    \enar
In this paper we restrict our attention to this subset of Horndeski theory.

Varying the action in Eq.~($\ref{horn}$) with respect to the metric~$g^{\mu\nu}$, and with respect to the scalar field $\phi$, we obtain the field equations and the Klein-Gordon-like equation, respectively
\begin{widetext}
    \begin{align}
        -\frac{1}{2} G_{2,X} \nabla_\mu \phi \nabla_v \phi-\frac{1}{2} G_2 g_{\mu \nu} +
        \frac{1}{2} G_{3,X} \square \phi \nabla_\mu \phi \nabla_v \phi+\nabla_{(\mu} G_3 \nabla_{v)} \phi-\frac{1}{2} g_{\mu \nu} \nabla_\lambda G_3 \nabla^\lambda \phi
        \nnn\label{EFE}
        +G_4 G_{\mu \nu} +g_{\mu v}\left(G_{4, \phi} \square \phi-2 X G_{4,\phi \phi}\right)-G_{4,\phi} \nabla_\mu \nabla_v \phi-G_{4,\phi \phi} \nabla_\mu \phi \nabla_v \phi &=\frac12 T_{\mu\nu}\,,
        \\[15pt]
        -\nabla_\mu G_{2,X} \nabla^\mu \phi-G_{2,X} \square \phi-G_{2,\phi}+2 G_{3, \phi} \square \phi+\nabla_\mu G_{3, \phi} \nabla^\mu \phi 
         +\nabla_\mu G_{3, X} \square \phi \nabla^\mu \phi
        \nnn\label{KG}
        +\nabla_\mu G_{3, X} \nabla^\mu X+G_{3, X}\left[(\square \phi)^2-\right. \left.\left(\nabla_\alpha \nabla_\beta \phi\right)^2-R_{\mu v} \nabla^\mu \phi \nabla^v \phi\right]-G_{4,\phi} R&=0\,,
    \end{align}
\end{widetext}
where the energy-momentum tensor~$T_{\mu\nu}$ is associated with the matter Lagrangian.
Independently of the gravity sector, the energy-momentum tensor is covariantly conserved as a consequence of diffeomorphism symmetry
    \bear\label{E-M}
    \nabla_\mu T^{\mu\nu}=0\,.
    \enar

The phenomenology of Horndeski gravity is very rich (see, e.g.,~\cite{cliftonModifiedGravityCosmology2012,heisenbergSystematicApproachGeneralisations2019,koyamaCosmologicalTestsModified2016,creminelliDarkEnergyInstabilitiesInduced2020,defeliceTheories2010,gleyzesHealthyTheoriesHorndeski2015,gleyzesEssentialBuildingBlocks2013,charmousisGeneralSecondOrder2012,sotiriouTheoriesGravity2010,copelandDynamicsDarkEnergy2006a,heisenbergHorndeskiQuantumLoupe2020}), and for some given choices of free functions it describes well-known classes of scalar-tensor theories. For example, Brans-Dicke theory~\cite{bransMachPrincipleRelativistic1961} is obtained in the limit
    \bear\label{BDlimit}
    G_2(\phi,X)\to \frac\omega\phi X\,,\quad G_3(\phi,X)\to0\,,\quad
    G_4(\phi)\to\frac12\phi \,.\quad
    \enar
Finally, general relativity is recovered for $G_4=(16\pi G_N)^{-1}$ with vanishing~$G_2$ and~$G_3$ in Eq.~($\ref{horn}$).

The simplest cosmological solution governed by Horndeski theory is the evolution of a spatially flat FLRW universe specified by the background metric
    \bear\label{FRW}
    ds^2=-a^2(\eta)d\eta^2+a^2(\eta)\de_{ij}dx^i dx^j\,.
    \enar
Given the high degree of symmetry of the homogeneous and isotropic FLRW metric, the only consistent scalar field configuration is a time-dependent function~$\bar\phi(\eta)$, and the matter content is described by the total background energy density~$\bar\rho(\eta)$ and pressure~$\bar p(\eta)$. A straightforward substitution of the FLRW quantities in the field equations~($\ref{EFE}$) leads to the generalization of the Friedmann equations in general relativity, while the same substitution in Eq.~($\ref{KG}$) leads to the background Klein-Gordon equation:
\begin{widetext}
	\bear\label{bk1}
        a^2\bar\rho&=&\frac12a^2G_2+3\HH^2 G_4-\frac12\bar\phi'^2 G_{2,X}-\frac3{2a^2}\HH\bar\phi'^3 G_{3,X}+\frac12\bar\phi'^2 G_{3,\phi}+3\HH\bar\phi'G_{4,\phi}\,,
	\nnn
        a^2\bar p&=&-\frac12a^2G_2-\left(\HH^2+2\HH'\right)G_4+\frac1{2a^2}\bar\phi'^2\left(\bar\phi''-\bar\phi'\right) G_{3,X}+\frac12\bar\phi'^2 G_{3,\phi}
	-\left(\bar\phi''+\HH\bar\phi' \right)G_{4,\phi}-\bar\phi'^2 G_{4,\phi\phi}\,,\quad
        \\[15pt]
        a^4\left(3\bar p-\bar\rho \right)&=&
    -a^2\left(\bar\phi''+2\HH\bar\phi' \right)G_4G_{2,X}-\bar\phi'^2\left(\bar\phi''-\HH\bar\phi' \right)G_4G_{2,XX}+a^4G_4G_{2,\phi}-a^2\bar\phi'^2G_4G_{2,\phi X}
    \nnn&&
    -3\bar\phi'\left(2\HH\bar\phi''+\HH'\bar\phi' \right)G_4G_{3,X}-\frac{3}{a^2}\HH\bar\phi'^3\left(\bar\phi''-\HH\bar\phi' \right)G_4G_{3,XX} +2a^2 \left(\bar\phi''+2\HH\bar\phi' \right)G_4 G_{3,\phi}
    \nnn&&
    +\bar\phi'^2\left(\bar\phi''-4\HH\bar\phi' \right) G_4G_{3,\phi X}-2a^4G_2G_{4,\phi}+\frac12a^2\bar\phi'^2G_{2,X}G_{4,\phi}+\frac32\bar\phi'^2\bar\phi''G_{3,X}G_{4,\phi}
    \nnn&&
    +a^2\bar\phi'^2G_{3,\phi}G_{4,\phi}-3a^2\left(\bar\phi''+2\HH\bar\phi'\right)G^2_{4,\phi} +a^2\bar\phi'^2 G_4G_{3,\phi\phi}-3a^2\bar\phi'^2G_{4,\phi}G_{4,\phi\phi}\,.
    \enar
\end{widetext}
where a prime indicates derivatives with respect to the conformal time~$\eta$, and~$\HH=\frac{a'}a$ is the conformal Hubble parameter.

The conservation of the energy-momentum tensor in Eq.~($\ref{E-M}$) reduces to
    \bear
    \bar\rho'=-3\HH(\bar p+\bar\rho)\,,
    \enar
which is the same relation as in general relativity.

	\section{Infrared Solutions of Horndeski Theory}\label{due}
 
In this Section we derive the linear-order Horndeski field equations in the uniform scalar-field gauge and investigate for infrared solutions. To simplify the complicated equations while preserving the physical picture, we resort to the Brans-Dicke limit of Horndeski gravity to find infrared solutions and use them to infer the solutions of Horndeski theory.

	\subsection{Linear-order evolution equations in Horndeski cosmology}\label{linpert}

At linear order in perturbations the line element is given in Eq.~($\ref{FRWpert}$), and the scalar field~$\phi(x)$ can be split as~$\bar\phi(\eta)+\de\phi(\eta,\bm x)$.
The covariant conservation of the energy-momentum tensor in Eq.~($\ref{E-M}$) at linear order becomes
    \bear\label{cons0}
    \de\rho'+3\HH(\de p+\de\rho)+3\left(\bar p+\bar\rho\right)\varphi'=\left(\bar p+\bar\rho\right)\Delta v\,,
    \enar
for the time component, and
    \bear\label{consi}
    \pa_i\left(\al+\frac{\de p}{\bar p+\bar\rho}-\frac{1}{a^4\left(\bar p+\bar\rho\right)}
    \pa_\eta\left[a^4\left(\bar p+\bar\rho\right)v\right]\right)=0\,,
    \enar
for the space component. Both equations are valid in any choice of gauge condition, and again they are independent of gravity theories in Jordan frames.\footnote{The expressions for the observables in the Jordan and Einstein frames coincide because the conformal transformation that relates the two frames does not affect light propagation. However, since the energy-momentum tensor is not covariantly conserved in the Einstein frame, the conditions for infrared insensitivity will take a different form in the Einstein frame.}

To facilitate the computation of the linear-order Horndeski field equations we exploit the gauge freedom in the diffeomorphism symmetry and choose the uniform scalar-field gauge defined by
    \bear\label{gauge}
    \phi_u\equiv \bar\phi(\eta) \Dquad \de\phi_u\equiv0\,,
    \enar
as the temporal gauge condition, and~$\ga\equiv0$ as our spatial gauge condition, so that there is no remaining gauge freedom.
Expanding Eq.~($\ref{EFE}$) to linear order in perturbations, and taking the uniform scalar-field gauge we derive
\begin{widetext}
    \bear\label{00h}
    \frac{a^2}2\de\rho_u&=&\bigg[ -6\HH\left(\HH G_4+\bar\phi' G_{4,\phi}\right)+\frac{\bar\phi'^4}{2a^2}\left( G_{2,XX}-G_{3,\phi X}+\frac{12}{\bar\phi'}\HH G_{3,X}+\frac3{a^2}\HH\bar\phi'G_{3,XX} \right)  
    \nnn&&
    +\bar\phi'^2\left(\frac{G_{2,X}}2-G_{3,\phi}\right)\bigg]\alpha_u
    +3\left(2\HH G_4-\frac{\bar\phi'^3G_{3,X}}{2a^2}+\bar\phi'G_{4,\phi}\right)\left(\varphi'_u+\frac{\Delta\be_u}{3a}\right)
    -2G_4\Delta\varphi_u\,,
    \\[15pt]
    \label{0ih}
    \frac{a^2}2\left(\bar p+\bar\rho\right)\pa_iv_u&=&\left(2\HH G_4-\frac{\bar\phi'^3 G_{3,X}}{2a^2}+\bar\phi' G_{4,\phi}\right)\pa_i\al_u-2G_4\pa_i\varphi'_u\,,
    \\[15pt]
    \label{ijh}
    0&=&\pa_i\pa_j\left[G_4\left(\frac{\be_u}{a}\right)'+\left(\frac{\be_u}a\right)\left( 2\HH G_4+\bar\phi'G_{4,\phi}\right)-G_4\left(\al_u+\varphi_u \right)\right]\,,
    \\[15pt]
    \label{iih}
    \frac{a^2}2\de p_u&=&\bigg[\left(\HH^2+2\HH'\right)G_4+\frac14\bar\phi'^2G_{2,X}-\frac{\bar\phi'^2}{a^2}\left(\bar\phi''-\HH\bar\phi' \right)G_{3,X}-\frac{\bar\phi'^4}{4a^2}\left(\bar\phi''-\HH\bar\phi' \right)G_{3,XX}-\frac12\bar\phi'^2G_{3,\phi}
    \nnn&&
    -\frac{\bar\phi'^4}{4a^2}G_{3,\phi X}+\left(\bar\phi''+\HH\bar\phi'\right)G_{4,\phi}+\bar\phi'^2G_{4,\phi\phi}\bigg]\al_u+\left(\HH G_4\frac1{4a^2}\bar\phi'^3G_{3,X}+\frac12\bar\phi'G_{4,\phi}\right)\al'_u
    \nnn&&
    -\left(2\HH G_4+\bar\phi' G_{4,\phi}\right)\left(\varphi'_u+\frac{\Delta\be_u}{3a}\right)-G_4\left[\varphi''_u+\left(\frac{\Delta\be_u}{3a}\right)'-\frac{\Delta\al_u}3-\frac{\Delta\varphi_u}3\right] \,,
    \enar
 \end{widetext}
which correspond respectively to the time-time component, the time-space components, the trace-less spatial components with $i\neq j$, and the spatial trace.

As our interest lies in the long-wavelength fluctuations, the Laplacians in the above equations can be ignored considering only the lowest-order coefficients~$f_0$ and~$f_1$ in our spatial expansion.
We remark that Eqs.~($\ref{consi}$) and~($\ref{0ih}$) are trivially satisfied by~$f_0$, while Eq.~($\ref{ijh}$) involves only orders higher than~$f_1$ due to the spatial derivatives applied to the equation. As noted by Weinberg \cite{weinbergAdiabaticModesCosmology2003}, since~$f_0$ and~$f_1$ need to be the large-scale limit of any physical solutions, the solutions to Eqs.~($\ref{consi}$),~($\ref{0ih}$), and~($\ref{ijh}$) should be continuous on all scales, so that we impose the equality without overall spatial derivatives.

        \subsection{Brans-Dicke limit of Horndeski theory}

A simple analytical solution of the background equations can be derived in the Brans-Dicke limit \cite{cliftonModifiedGravityCosmology2012}. Taking the limit in Eq.~($\ref{BDlimit}$) the background equations in Section~\ref{hornd} become
    \beeq\label{BDback}
        \begin{split}
        3\HH\bar\phi'+3\HH^2\bar\phi -\frac\omega2\frac{\bar\phi'^2}{\bar\phi}&=a^2\bar\rho\,,
	\\
        \bar\phi''+\HH\bar\phi'+\HH^2\bar\phi+2\HH'\bar\phi+\frac\omega2\frac{\bar\phi'^2}{\bar\phi}&=-a^2\bar p\,,
	\\
        \bar\phi''+2\HH\bar\phi'&=\frac{\bar\rho-3\bar p}{3+2\omega}a^2\,.
        \end{split}
    \eneq
Combining these equations we obtain the background governing equation for the scalar field
    \bear\label{master}
    \bar\phi''+2\HH\bar\phi'-\frac3\omega\left( \HH^2+\HH'\right)\bar\phi-\frac12\frac{\bar\phi'^2}{\bar\phi}=0\,,
    \enar
which does not contain matter fields directly. We look for power-law solutions of the form
    \bear\label{power}
    a(\eta)\propto\eta^n\,,\Dquad \bar\phi(\eta)\propto \eta^m\,,
    \enar
by assuming a constant equation of state~$\bar p=W\bar\rho$, and plugging the power-law ansatz in the first equation of~($\ref{BDback}$) we obtain
    \bear
    m=2-n(1+3W)\,.
    \enar
Finally, using the governing equation for the scalar field~($\ref{master}$) we derive the exponents
    \beeq
        \begin{split}
        n&=\frac{1+\omega\left(1-W\right)}{1+(\omega/2)(1+2W-3W^2)}\,,
        \\
        m&=\frac{1-3W}{1+(\omega/2)(1+2W-3W^2)}\,,        
        \end{split}
    \eneq
and the power-law solutions of the background equations in Brans-Dicke gravity are
    \beeq
        \begin{split}
        a(\eta)&=a_0\left(\frac{\eta}{\eta_0}\right)^{\frac{1+\omega\left(1-W\right)}{1+(\omega/2)(1+2W-3W^2)}}\,,
        \\
        \bar\phi(\eta)&=\frac1{8\pi G_N} \left(\frac{\eta}{\eta_0}\right)^{\frac{1-3W}{1+(\omega/2)(1+2W-3W^2)}} \,,
        \end{split}
    \eneq
where the proportionality constants are taken to match the general-relativity limit for~$\omega\to\infty$:
    \bear
    a(\eta)=a_0\left(\frac{\eta}{\eta_0}\right)^{\frac2{1+3W}}\,,\qquad \bar\phi(\eta)=\frac1{8\pi G_N}\,,
    \enar
where~$\eta_0$ is an arbitrary reference time at which the scale factor~is $a_0$, and~$G_N$ is the Newton gravitational constant.

Noteworthy limits of the background solutions are the case of matter-dominated era (MDE) where the universe is filled only by pressure-less matter, i.e.,~$W=0$
    \bear\label{mde}
    a(\eta)=a_0\left(\frac{\eta}{\eta_0}\right)^{\frac{1+\omega}{1+\omega/2}}\,,\qquad \bar\phi(\eta)=\frac1{8\pi G_N}\left(\frac{\eta}{\eta_0}\right)^{\frac{2}{2+\omega}},
    \enar		
and the case of radiation-dominated era (RDE) with~$W=1/3$
    \bear\label{RDE}
    a(\eta)=a_0\left(\frac{\eta}{\eta_0}\right)\,,\Dquad \bar\phi=\frac1{8\pi G_N}\,.
    \enar
The case of MDE contains nontrivial power-law solutions that indeed reduce to the FLRW solutions of general relativity in the limit~$\omega\to\infty$. On the other hand, the scaling of the RDE solution coincides with that of general relativity, without any dependence on the parameter~$\omega$ of the Brans-Dicke theory as the scalar field in RDE becomes non-dynamical.

Taking the limit in Eq.~($\ref{BDlimit}$) of the linear-order field equations of the Horndeski theory in Section~\ref{linpert}, we derive the linear-order evolution equations in Brans-Dicke theory in the uniform scalar-field gauge. The time-time component of the field equations reads
    \bear\label{prima}
    \al_u\left( \omega\frac{\bar\phi'^2}{\bar\phi} -6\HH\bar\phi'-6\HH^2\bar\phi\right)+3\varphi_u'\left( \bar\phi'+2\HH\bar\phi \right)=a^2\de\rho_u\,,~~
    \enar
while the time-space
    \bear\label{seconda}
    \pa_i\al_u\left(\bar\phi'+2\HH\bar\phi\right)-2\bar\phi\pa_i\varphi_u'=a^2\left(\bar\rho+\bar p\right)\pa_iv_u\,,
    \enar
the trace-less spatial components with~$i\neq j$
    \bear\label{resta}
    \pa_i\pa_j\left[\bar\phi\left(\frac{\be_u}a\right)'+\left(\frac{\be_u}a\right)\left(\bar\phi'+2\HH\bar\phi\right)-\bar\phi\left( \al_u+\varphi_u \right)\right]=0\,,\quad
    \enar
and the spatial trace
    \beeq\label{ultima}
        \begin{split}
         &\al_u\left( \bar\phi''+\HH\bar\phi'+\HH^2\bar\phi+2\HH'\bar\phi+\frac\omega2\frac{\bar\phi'^2}{\bar\phi} \right)-\bar\phi\varphi_u''
         \\&
        +\left(\bar\phi'+2\HH\bar\phi \right)\left(\al_u'-2\varphi_u' \right)
	=\frac{a^2}2\de p_u\,. 
        \end{split}
	\quad
    \eneq
Infrared solutions in the Brans-Dicke theory will be computed in Section~\ref{irhornd} together with infrared solutions in Horndeski theory.

	\subsection{Infrared sensitivity of cosmological probes in Horndeski theory}\label{irhornd}

Having derived the Horndeski field equations we are ready to address the issue of infrared sensitivity of the cosmological probes by studying the conditions in Section~\ref{ir}, i.e., the difference between the two curvature perturbations~$\RR$ and~$\zeta$ in Horndeski theory. From the definitions of the curvature perturbations in Eq.~($\ref{defcurv}$) a straightforward substitution yields
    \bear\label{diff}
    \RR-\zeta=-\frac{1}{3\left(\bar p +\bar\rho\right)}\left[\de\rho+3\HH\left(\bar p+\bar\rho\right)v \right]\,.
    \enar
We remark that the difference in the curvature perturbations is gauge invariant at linear order in perturbations, hence its computation is independent of our gauge choice.
Using the time-time and time-space components of the field equations~($\ref{00h}$) and~($\ref{0ih}$), we can rearrange the equation for~$\RR-\zeta$ in the uniform scalar-field gauge ($\de\phi_u\equiv0$) as
\begin{widetext}
    \bear\label{main}
    -3a^2\left(\bar p+\bar\rho\right)\left(\RR-\zeta\right)&=&\frac{\bar\phi'}{2}\bigg[ \frac{9}{a^2}\HH\bar\phi'^2 G_{3,X} +\frac{3}{a^4}\HH\bar\phi'^4 G_{3,XX} +\bar\phi'\left(G_{2,X}-2G_{3,\phi}\right)
    \nnn&&
    +\frac{\bar\phi'^3}{a^2}\left( G_{2,XX}-G_{3,\phi X} \right)-6\HH G_{4,\phi}\bigg]\al_u
    +\frac{\bar\phi'}2\left( 6 G_{4,\phi}-\frac{3}{a^2}\bar\phi'^2G_{3,X}\right)\varphi'_u\,,
    \enar
\end{widetext}
where the Laplacians in Eq.~($\ref{00h}$) have been neglected as we are interested in the large-scale limit, i.e., when only the two lowest-order coefficients in the spatial expansion in Eq.~($\ref{exp}$) are considered. 
Here we look for solutions to Eq.~($\ref{main}$) under the assumption that the matter content satisfies the adiabaticity condition in Eq.~($\ref{adiab}$) on large scales.

The covariant conservation of the energy-momentum tensor in Eqs.~($\ref{cons0}$) and~($\ref{consi}$), together with the adiabatic condition allow us to derive the following expressions that relate the metric perturbations to the fluctuations in the matter fields:
    \beeq\label{alphi}
        \begin{split}
        \varphi_u'=&-\frac{3\HH\left( \bar p'+\bar\rho'                \right)\de\rho_u+\bar\rho'\de\rho_u'}{3\left(\bar p +\bar\rho\right)\bar\rho'}\,,
        \\
	\alpha_u=&\left( -1+\frac{\HH'}{\HH^2}-\frac{\bar p'}{\HH(\bar     p+\bar\rho)} \right)\left(\RR-\zeta\right)-\frac{\RR'}{\HH}
        \\&
        +\frac1{\bar\rho'^2}\left[\left(\HH\bar\rho-\bar\rho''\right)\de\rho_u+\de\rho_u'\right]\,,
        \end{split}
    \eneq
where we used Eq.~($\ref{diff}$) to replace the velocity potential~$v$, and we evaluated the perturbations in the uniform scalar-field gauge. With such results we can recast Eq.~($\ref{main}$) to the simple form
    \bear\label{replace}
    \CC_1\left(\RR-\zeta\right)+\CC_2\de\rho_u+\CC_3\de\rho_u'+\CC_4\RR'=0\,,
    \enar
where we defined the following background functions of conformal time:
    \bear
    \CC_1&=&\HH\left(-1+\frac{\HH'}{\HH^2}-\frac{\bar p'}{\HH(\bar p+\bar\rho)} \right)\CC_4-3a^2\left(\bar p +\bar\rho\right)\,,
    \\
    \CC_2&=&
    -\frac{\bar\phi'}{2a^2(\bar p+\bar\rho)}\left(\frac{\bar p'+\bar\rho'}{\bar p+\bar\rho}\right)\left(\bar\phi'^2 G_{3,X}-2a^2G_{4,\phi}\right)
    \nnn&&
    +\HH\left(\frac{\HH}{\bar\rho'}-\frac{\bar\rho''}{\bar\rho'^2} \right)\CC_4\,,
    \\
    \CC_3&=&\frac{\HH}{\bar\rho'}\CC_4-\frac{\bar\phi'}{2a^2(\bar p+\bar\rho)}\left(\bar\phi'^2 G_{3,X}-2a^2G_{4,\phi}\right)\,,
    \\
    \CC_4&=&\frac{\bar\phi'}{2\HH}\bigg[ \frac{9}{a^2}\HH\bar\phi'^2 G_{3,X} +\frac{3}{a^4}\HH\bar\phi'^4 G_{3,XX} +\bar\phi'\big(G_{2,X}
    \nnn&&
    -2G_{3,\phi}\big)+\frac{\bar\phi'^3}{a^2}\left( G_{2,XX}-G_{3,\phi X} \right)-6\HH G_{4,\phi}\bigg]\,.\qquad
    \enar
Thanks to the simple structure of the ordinary differential equation in Eq.~($\ref{replace}$) it is evident that an infrared solution with~$\RR=\zeta$ in Horndeski theory is guaranteed if~$\de\rho_u=0$. We discuss this solution later, and first check if other solutions with~$\RR=\zeta$ but~$\de\rho_u\neq0$ exist. In particular, since~$\zeta'=0$, we look for more general solutions with~$\RR'=0$.

As an explicit example, we take the Brans-Dicke limit in Eq.~($\ref{BDlimit}$) and consider a universe filled with pressure-less matter. Using the background solution for MDE in Eq.~($\ref{mde}$), the governing equation~($\ref{replace}$) simplifies to
    \beeq	
        \begin{split}\label{BDMDE}
        &\frac{1}{\bar\rho_0}\left(\frac{\eta}{\eta_0}\right)^{\frac{6+6\omega}{2+\omega}}\bigg[(12+11\omega)\de\rho_u-\omega(2+\omega)\eta\de\rho_u'\bigg]
        \\&
        +3\left(3+2\omega\right)\left(4+3\omega\right)^2\left(\RR_m-\zeta_m\right)
        =0\,,\qquad
	\end{split}
    \eneq
where~$\bar\rho_0$ is the background density value at the reference time~$\eta_0$.
With a power-law ansatz for the density fluctuation in the uniform scalar-field gauge
    \bear
    \de\rho_u=\bar\rho_0\de_u\left(\frac{\eta}{\eta_0}\right)^r\,,
    \enar
we derive two solutions to Eq.~($\ref{BDMDE}$):~$\RR'_m=\zeta'_m=0$ but~$\RR_m\neq\zeta_m$,
    \bear\label{nonadiab}
    \de\rho_u=-3\bar\rho_0(4+3\omega)\left(\RR_m-\zeta_m\right)\left(\frac{\eta}{\eta_0}\right)^{-\frac{6+6\omega}{2+\omega}}\,,
    \enar
and one for which~$\RR'_m=\zeta'_m=0$ and~$\RR_m=\zeta_m$,
    \bear\label{BDMDErho}
    \de\rho_u= \bar\rho_0\de_u\left(\frac{\eta}{\eta_0}\right)^{\frac{12+11\omega}{\omega(2+\omega)}} \,,
    \enar
where the constant amplitude~$\de_u$ is undetermined.
Note, however, that as we have only used the time-time and time-space components of the field equations, together with the conservation of the energy-momentum tensor, we need to verify if these solutions also satisfy the spatial trace component of the field equations.
In Brans-Dicke gravity the field equation for the spatial trace is given in Eq.~($\ref{ultima}$), and it simplifies in MDE as
    \bear
    &&6(10+19\omega+9\omega^2)\de\rho_u+2(2+\omega)(9+8\omega)\left(\frac{\eta}{\eta_0}\right)\de\rho'_u
    \nnn&&
    +(2+\omega)^2\left(\frac{\eta}{\eta_0}\right)^2\de\rho''_u=0\,,    
    \enar
where we used Eq.~($\ref{alphi}$) again and looked for solution with~$\RR'_m=0$. A general solution to this second-order ordinary differential equation is given in terms of power laws:
    \bear
    \de\rho_u=\bar\rho_0\de_{u_1} \left(\frac{\eta}{\eta_0}\right)^{-\frac{10+9\omega}{2+\omega}}+\bar\rho_0\de_{u_2} \left(\frac{\eta}{\eta_0}\right)^{-\frac{6+6\omega}{2+\omega}}\,.
    \enar
The only non-trivial solution with $\de\rho_u\neq0$ that is consistent with the field equations is the one in Eq.~($\ref{nonadiab}$) with~$\RR_m\neq\zeta_m$, for which the cosmological probes are sensitive to the presence of infrared fluctuations.

Although our computation has been carried out only for the simple limit of Brans-Dicke gravity in MDE, the conclusion is also applicable to the more general Horndeski theory, but of course with more involved algebra. For instance, a non-trivial solution to Eq.~($\ref{replace}$) with~$\RR=\zeta$ implies the following profile for the density fluctuation in the uniform-field gauge:
    \bear
    \de\rho_u=\bar\rho_0\de_u \exp\left({-\int_{\eta_0}^\eta d\tilde\eta\,\frac{\CC_2}{\CC_3}}\right)\,,
    \enar
which can then be substituted in the spatial trace component in Eq.~($\ref{iih}$). With the adiabatic condition on large scales and by employing Eq.~($\ref{alphi}$) the resulting ordinary differential equation takes the functional form
    \bear\label{implicit}
    F[G_i,\bar\phi,a,\bar p,\bar\rho]\de_u=0\,,
    \enar
where~$F[G_i,\bar\phi,a,\bar p,\bar\rho]$ is an involved expression that contains derivatives of the background Horndeski quantities. Given the unique time dependence of the background quantities, the function~$F$ is a non-trivial function of time. Therefore, the only solution with~$\RR=\zeta$ to Eq.~($\ref{implicit}$) is one with~$\de\rho_u=0$ on large scales. In conclusion, there is only one infrared solution in Horndeski gravity, for which the cosmological probes are insensitive to the presence of infrared fluctuations. This solution is characterized by
    \beeq\label{IRinsol}
        \begin{split}
        \varphi_u\equiv \RR=\zeta\,,\Dquad \be_u=\frac{\RR}{a G_4}\int_0^\eta d\tilde\eta \,a^2 G_4 \,, 
        \\
        \al_u=v_u=\de\rho_u=\de p_u=0\Dquad\,,
        \end{split}
    \eneq
where~$\RR$ is set by the initial conditions and the solution for~$\be_u$ is obtained integrating Eq.~($\ref{ijh}$). Importantly, the solution in general relativity without infrared sensitivity has the same structure, with the only difference that~$G_4$ is replaced by~$(16\pi G_N)^{-1}$, which is indeed true in the general-relativity limit of the Horndeski theory.

	\section{Adiabatic Modes in Horndeski Theory}\label{tre}

Here we attempt to gain a better understanding of the two solutions in Horndeski theory, with and without infrared sensitivity of cosmological probes by connecting them to diffeomorphism symmetry in the infrared limit, i.e., adiabatic modes à la Weinberg~\cite{weinbergAdiabaticModesCosmology2003}.
Given a general but infinitesimal coordinate transformation $\tilde x^\mu(x)=x^\mu+\xi^\mu(x)$, the scalar field fluctuation gauge transforms as~\cite{bardeenGaugeinvariantCosmologicalPerturbations1980,mukhanovTheoryCosmologicalPerturbations1992a}.
    \bear\label{scalar}
    \widetilde{\de\phi}=\de\phi_u-\xi^0\bar\phi'= -\xi^0\bar\phi'\,.
    \enar
and the fluid quantities transform as
    \beeq\label{fluid}
        \begin{split}
        \widetilde{\de\rho}=\de\rho_u-\xi^0\bar\rho'\,,\Dquad \widetilde{\de p}=\de p_u-\xi^0\bar p'\,, 
        \\
        \pa_i\widetilde v=\pa_i v_u -\pa_i \xi^0 \,.\Dquad
	\end{split}
    \eneq
The metric perturbations in Eq.~($\ref{FRWpert}$) also gauge transform as
    \bear\label{metric}
    \widetilde\varphi\de_{ij}+\pa_i\pa_j\widetilde\ga=\varphi_u\de_{ij}-\HH\xi^0\de_{ij}-\frac12\pa_i\xi^k\de_{kj}-\frac12\pa_j\xi^k\de_{ki}\,,
    \nnn
    \widetilde\al=\,\al_u-\frac1a\left(a\xi^0\right)' \,,\qquad \pa_i\widetilde\be=\pa_i\be_u-\pa_i\xi^0 +\xi'^j\de_{ij}\,,\quad
    \enar
where~$\ga\equiv0$ in the uniform scalar-field gauge. If the gauge vector~$\xi^\mu$ goes to zero at spatial infinity, which is the ordinary case (or \textit{small} gauge transformations), then there exists no residual gauge freedom in the uniform scalar-field gauge. However, if $\xi^\mu$ induces instead a \textit{large} gauge transformation that is not bounded in space, the uniform scalar-field gauge has residual gauge freedom in the infrared limit~\cite{mitsouLargeGaugeTransformations2022}.

We exploit this redundancy to build physical solutions from gauge modes as shown in Weinberg~\cite{weinbergAdiabaticModesCosmology2003}.
Consider a background solution of Horndeski gravity with \textit{no} perturbations, and perform a coordinate transformation. The gauge transformation equations~($\ref{scalar}$),~($\ref{fluid}$), and~($\ref{metric}$) can be used to show how gauge modes are generated as a consequence of the coordinate transformation of the background solution without any perturbations:
    \beeq
	\begin{split}
	\frac{\de\phi}{\bar\phi'}&=\frac{\de\rho}{\bar\rho'}=\frac{\de p}{\bar p'}=-\xi^0\,, \qquad \pa_iv=-\pa_i\xi^0\,,
	\\
	\al&=-\frac1a\left(a\xi^0\right)' \,,\qquad\quad\pa_i\be=-\pa_i\xi^0 +\xi'^j\de_{ij}\,, \quad
 \\&\qquad
 \varphi\de_{ij}+\pa_i\pa_j\ga=-\HH\xi^0\de_{ij}-\pa_{(i}\xi_{j)}\,.
	\end{split}
    \eneq
These perturbation solutions are equivalent to the background solution, which can be removed again, hence they represent the residual gauge mode in the infrared limit of vanishing Fourier mode~$k=0$. With our choice of the uniform scalar-field gauge condition in Eq.~($\ref{gauge}$), the temporal part of the transformation is fixed~$\xi^0=0$, and the matter content as well as the scalar field satisfies the adiabatic condition.
Moreover, the spatial gauge condition~$\ga\equiv0$ constrains the expression for the curvature perturbation~$\varphi_u$ to be related to spatial part~$\xi^i$ of the transformation in a specific way. Considering a large gauge transformation of the form $\xi^\mu(x)=\cc(\eta)x^i\de^\mu_i$ (see \cite{pajerSystematicsAdiabaticModes2018} for more sophisticated choices), the only non-vanishing fictitious perturbations in the uniform scalar-field gauge conditions are
    \bear\label{gaugemodes}
    \varphi_u\equiv R=\zeta=-\frac13\lambda\,,\qquad \be_u=f(\eta)+\frac12\lambda' x^ix^j\de_{ij}\,.
    \enar
Though this solution is a pure gauge mode at zero Fourier mode~$k=0$, it is possible to promote it to a physical solution by demanding that the perturbation solutions in Eq.~($\ref{gaugemodes}$) satisfy the field equations in Eqs.~($\ref{0ih}$) and~($\ref{ijh}$) at~$k>0$. The physical solution is
    \bear\label{Weinb}
    \lambda'=0\,,\Dquad f(\eta)= -\frac{\cc}{3a G_4}\int_0^\eta d\tilde\eta \,a^2 G_4\,,
    \enar
and it is the adiabatic mode à la Weinberg in Horndeski theory.

Remarkably, we notice that the solution as an adiabatic mode is equivalent to the solution in Section~\ref{irhornd} [Eq.~($\ref{IRinsol}$)] without infrared sensitivity in the cosmological probes. It is now clear that the physical condition for such a solution is the generalized adiabaticity among all matter content including the Horndeski scalar field:\footnote{In this generalized sense we consider the Horndeski scalar field~$\phi$ as part of the matter content, though it is actually contributing to the gravity sector and does not have an energy-momentum tensor with simple fluid description.}
    \bear\label{Wadiab}
    \frac{\de p_u}{\bar p'}=\frac{\de\rho_u}{\bar\rho'}=\frac{\de\phi_u}{\bar\phi'}=0\,.
    \enar
Notice that the last equality is specific to the uniform scalar-field gauge, while the other equalities are valid in any gauge. The other infrared solution in Section~\ref{irhornd} [Eq.~($\ref{nonadiab}$)] with infrared sensitivity ($\RR'=0$, $\RR\neq\zeta$) is in fact not adiabatic in this generalized sense:
    \bear
    \frac{\de\rho_u}{\bar\rho'_m}=\frac{(2+\omega)(4+3\omega)}{2+2\omega}\left(\RR_m-\zeta_m\right)\eta\neq\frac{\de\phi_u}{\bar\phi'_m}=0\,.
    \enar
We conclude that if all matter fields satisfy the generalized adiabatic condition, the general covariance of the theory ensures the existence of a long-wavelength adiabatic solution, which is continuously mapped into a pure gauge mode at $k=0$ that can be removed by a coordinate transformation, i.e., the presence of such infrared fluctuations cannot affect cosmological probes. Given that this conclusion is valid in general relativity beyond linear-order perturbations~\cite{mitsouInfraredSensitivityRelativistic2023}, we suspect that our conclusion in Horndeski theory is likely to remain valid non-linearly.

	\section{Summary and Discussion}\label{quattro}

Cosmological observations in large-scale surveys are used to construct cosmological probes such as galaxy clustering, weak gravitational lensing, luminosity distance, and cosmic microwave background anisotropies. Theoretical descriptions of the cosmological probes show that they can be constructed with a minimal assumption that the light propagation follows null geodesics in a four-dimensional metric theory, independently of the underlying theory of gravity. Furthermore, their relativistic descriptions revealed that cosmological probes are affected by numerous relativistic effects~\cite{didioGalaxyNumberCounts2014,bertaccaObservedGalaxyNumber2014,koyamaObservedGalaxyBispectrum2018,jeongLargescaleClusteringGalaxies2012,bonvinWhatGalaxySurveys2011a,challinorLinearPowerSpectrum2011a,yooNewPerspectiveGalaxy2009,yooGeneralRelativisticDescription2010,yooLinearOrderRelativisticEffect2014,yooRelativisticEffectGalaxy2014,magiSecondorderGaugeinvariantFormalism2022}. However, the presence of relativistic effects in cosmological probes such as gravitational potential can be problematic, because a change in the uniform gravitational potential cannot have a physical impact on local measurements according to the equivalence principle. Applied to cosmological probes, any long-wavelength fluctuations can modulate the cosmological probes, but fluctuations with wavelength larger than the separation of the observer and the source should have a progressively diminishing impact as the wavelength becomes larger. This is indeed the case in general relativity --- long wavelength or infrared fluctuations exist in cosmological probes, but they add up to cancel each other only if the Einstein equations are used.

Motivated by this observation, we have derived~\cite{magiConditionsAbsenceInfrared2023a} the general conditions without using the Einstein equations, under which cosmological probes are devoid of such infrared sensitivity. We have shown in this work that the general conditions can be rephrased as the adiabatic conditions for the matter content and the equality of the comoving gauge curvature perturbation~$\RR$ and the uniform-density gauge curvature perturbation~$\zeta$. In general relativity, the equality is guaranteed by the Einstein equations in the infrared, and the adiabatic condition imposes that~$\zeta$ is constant in the infrared. To go beyond Einstein, we have investigated the infrared solutions first in the Brans-Dicke theory and then in the Horndeski theory. Despite the adiabaticity in the matter content, there exist solutions in the infrared where the general condition~$\RR=\zeta$ is violated and hence retain the sensitivity to infrared fluctuations. This class of solutions in Horndeski theory could lead to extraordinary phenomena such as modulation by super-horizon fluctuations or even pathology like infrared divergence, if infrared fluctuations are scale-invariant to the limit~$k=0$.

Horndeski theory with the Brans-Dicke theory included admits general relativity as a limiting case, and hence solutions that satisfy the general conditions are expected to exist in Horndeski theory. Indeed, we have found an explicit solution in the Brans-Dicke theory and thereby inferred a solution in Horndeski theory. This class of solutions in Eq.~($\ref{IRinsol}$) takes the same form as in general relativity, in which all the metric and fluid perturbations vanish in the infrared, except for
    \beeq
    \RR=\zeta\neq0~,\Dquad \be_u=\frac{\RR}{a G_4}\int_0^\eta d\tilde\eta \,a^2 G_4\,,
    \eneq
completely set by the non-vanishing constant~${\cal R}$ from the initial conditions. The Horndeski function~$G_4$ reduces to a constant in general relativity. Given the initial conditions in terms of~${\cal R}$, the value of~$\beta_u$ in the infrared can be different in Horndeski theory from general relativity, but the structure of perturbations is identical and the value of~$\beta_u$ is only affected by the background evolution. In short, the infrared sensitivity of cosmological probes, which could lead to pathological behavior in cosmological observations, can be avoided in Horndeski theory, for an infrared solution with the same structure of general relativity but with different numerical values.

To better understand the solution, we have considered adiabatic modes à la Weinberg~\cite{weinbergAdiabaticModesCosmology2003}. With diffeomorphism symmetry in general relativity and Horndeski theory, non-trivial solutions can be generated in the limit~$k=0$ by a large gauge transformation. Ordinary coordinate transformations, which can be obtained by a small deformation from identity, induce the standard (or small) gauge transformation, and perturbation variables are fully specified once the coordinate transformation is fixed (see, e.g., \cite{bardeenGaugeinvariantCosmologicalPerturbations1980}). In contrast, large gauge transformations, which are by definition {\it not} small gauge transformations, allow extra gauge freedom for perturbation variables at~$k=0$, even though they are completely specified at~$k>0$. Not all the perturbations generated by large gauge transformations are physical, rather fictitious gauge modes. However, some of those gauge modes can be smoothly connected to physical solutions at~$k>0$, and they are called the adiabatic modes. With a small gauge transformation fixed, a large gauge transformation applied to the background solution induces the fluctuations at~$k=0$ that satisfy the adiabatic condition.

Applied to Horndeski theory, we have found a physical adiabatic mode by a large gauge transformation (see also~\cite{crisostomiConsistencyRelationsLargescale2020, lewandowskiViolationConsistencyRelations2020} for a derivation of the adiabatic mode in Horndeski theory in the non-relativistic limit). The adiabatic mode that is valid at~$k>0$ takes the form of the solution we found by explicitly solving the field equation in the infrared, under the condition that cosmological probes are devoid of any infrared sensitivity. In other words, the general conditions, under which cosmological probes are not sensitive to infrared modes, are indistinguishable from a physical adiabatic mode that is continuously mapped into a pure gauge mode at~$k=0$. This gauge mode at~$k=0$ can be eliminated by a large gauge transformation, and hence the physical solution is equivalent to the situation where there is {\it no} infrared fluctuations and hence no impact on cosmological probes. Furthermore, the adiabatic mode from a large gauge transformation reveals that the adiabatic condition is imposed not only on the matter content, but also on the Horndeski scalar field. With our choice of a uniform scalar-field gauge, the adiabatic condition corresponds to~$\delta \rho_u=0$ in the infrared. Put it differently, the generalized adiabatic condition on the matter content, including the Horndeski scalar field, guarantees an infrared-insensitive solution with which no pathological behaviors exist in the cosmological probes, as in general relativity.

If the generalized adiabatic condition is violated in the infrared, cosmological probes in large-scale surveys can be sensitive to infrared fluctuations or exhibit pathological behavior, depending on the level of violation and the amplitude of infrared fluctuations. Such violation of the adiabatic condition is often found in multi-field inflationary models~\cite{mollerachIsocurvatureBaryonPerturbations1990,bartoloAdiabaticIsocurvaturePerturbations2001, byrnesCurvatureIsocurvaturePerturbations2006, gordonAdiabaticEntropyPerturbations2000, petersonTestingTwoFieldInflation2011, polarskiSpectraPerturbationsProduced1992} or cosmological defect models such as cosmic strings and branes~\cite{brandenbergerSearchingCosmicStrings2014, brandenbergerTopologicalDefectsStructure1994}. These models are often classified as models of isocurvature perturbations. Naturally, the presence of isocurvature perturbations violates the adiabatic condition, and the deviation in multi-fluid fluctuations in the infrared would lead to an apparent violation of the equivalence principle. Such violation can also arise from long-range non-gravitational interactions, for example, primordial non-Gaussianity in the initial condition~\cite{bartoloNonGaussianityInflationTheory2004, komatsuAcousticSignaturesPrimary2001}, interactions in the dark sector~\cite{archidiaconoUnveilingDarkFifth2022}, and large-scale cosmological back-reaction~\cite{buchertAveragePropertiesInhomogeneous2000, buchertAveragingInhomogeneousNewtonian1997, clarksonDoesGrowthStructure2011, comeauBackReactionLongWavelengthCosmological2023, geshnizjaniBackReactionLocal2002, marozziCosmologicalBackreactionTest2013}. Given the tight constraint on the level of isocurvature perturbations in CMB observations~\cite{aghanimPlanck2018Results2020, akramiPlanck2018Results2020a}, the level of infrared sensitivity in cosmological probes is small in general relativity (indeed exactly zero in $\Lambda$CDM model), and our work shows that this remains true in Horndeski theory.

What physical mechanism establishes the adiabatic condition in the infrared? In the standard model, an inflationary mechanism generates adiabatic fluctuations in the infrared, originating from a single inflaton field (see, e.g.,~\cite{brandenbergerQuantumFieldTheory1985,lindeInflationaryCosmology2000,lythParticlePhysicsModels1999}). While multi-field models can generate non-adiabatic perturbations in the infrared, these perturbations are often converted into adiabatic perturbations~\cite{mollerachPrimordialOriginIsocurvature1990,kodamaEvolutionIsocurvaturePerturbations1986,kodamaEvolutionIsocurvaturePerturbations1987,weinbergMustCosmologicalPerturbations2004}. Furthermore, any non-adiabatic perturbations in the initial conditions decay away in the infrared~\cite{weinbergCanNonadiabaticPerturbations2004} if a local thermal equilibrium is established after the perturbation generation. It appears that it is difficult in general relativity to maintain any deviation from the adiabatic condition in the infrared. What about the generalized adiabatic condition in theories beyond general relativity? As opposed to the $\Lambda$CDM model with a single source of perturbation generation, no physical mechanism is known to exist in the presence of an additional gravitational degree of freedom, which sets the same time shift for matter and gravitational scalar field.
Since infrared-insensitive solutions exist in both general relativity and Horndeski theory, we cannot use the existence of this type of solution as a criterion for preferring one theory over the other. However, given the simplicity of the mechanism by which the initial conditions for infrared-insensitive solutions are realized, we can conclude that general relativity is favored.

        \acknowledgments
        
We thank Robert Brandenberger, Ermis Mitsou, and Marko Simonović for useful discussions. We acknowledge support by the Swiss National Science Foundation. J.Y. is further supported by
a Consolidator Grant of the European Research Council.

\bibliography{biblio}

\end{document}